\documentclass[aps,prl,notitlepage,reprint,superscriptaddress,nofootinbib,longbibliography]{revtex4-2}

\usepackage[english]{babel}
\makeatletter
\let\l@en\l@english
\makeatother

\usepackage{booktabs}

\usepackage{graphicx}
\usepackage{xcolor}
\usepackage{dcolumn}
\usepackage{bm}
\usepackage{amsmath}
\usepackage{upgreek}
\usepackage[colorlinks,linkcolor=red,citecolor=blue,urlcolor=blue]{hyperref}
\usepackage[utf8]{inputenc}
\usepackage[T1]{fontenc}
\usepackage{mathptmx}
\usepackage{ulem}
\usepackage[colorinlistoftodos]{todonotes}
\usepackage[mathlines]{lineno}% Enable numbering of text and display math
\renewcommand{\t}[1]{\mathrm{#1}}
\begin{document}
	
	\title{Angular displacement readout of a mechanical oscillator with a guided mode resonance}
	
	\author{D. Torres-Barajas}
	\email{dtorresb@arizona.edu}
	\affiliation{Wyant College of Optical Sciences, University of Arizona, Tucson, AZ 85721, USA}

    \author{O. E. Angulo}
	\affiliation{Wyant College of Optical Sciences, University of Arizona, Tucson, AZ 85721, USA}

	\author{A. R. Agrawal}
	\affiliation{Wyant College of Optical Sciences, University of Arizona, Tucson, AZ 85721, USA}

  %  \author{A. D. Hyatt}
%	\affiliation{Wyant College of Optical Sciences, University of Arizona, Tucson, AZ 85721, USA}
    
	\author{M. ElKabbash}
    \email{melkabbash@arizona.edu}
	\affiliation{Wyant College of Optical Sciences, University of Arizona, Tucson, AZ 85721, USA}

    \author{D. J. Wilson}
	\email{dalziel@arizona.edu}
	\affiliation{Wyant College of Optical Sciences, University of Arizona, Tucson, AZ 85721, USA}
	
	\date{\today}
	\begin{abstract}
%Measuring the angular displacement of a mechanical oscillator is a ubiquitous task commonly performed by optical beam tracking; however, the free-space, multimode nature of this approach makes chipscale integration and coherent signal enhancement challenging. Here we demonstrate coherently-enhanced angular displacement readout with a guided mode resonance (GMR), applying it to precision readout of a nanomechanical oscillator. Specifically, we fabricate subwavelength gratings into 100-nm-thick Si$_3$N$_4$ membranes and record their vibration by direct transmission measurements.   The narrow GMR linewidth $\approx 2\;\text{mrad}$ enables a shot-noise-limited displacement imprecision of $10^{-9}\;\t{rad/\sqrt{Hz}}$ with nanowatts of optical power, sufficient to resolve the thermal motion of a $Q\approx 10^6$ torsion mode with a signal-to-noise ratio of 44 dB. Control experiments based on polarization and wavelength detuning confirm that the signal arises from GMR-mediated transduction. These results establish GMR as an on-chip approach to angular displacement readout in quantum optomechanical sensors.

 Measuring the angular displacement of a mechanical oscillator is a ubiquitous task; however, the multimode nature of angular optomechanical coupling makes coherent signal enhancement challenging. Here we demonstrate coherently enhanced angular displacement readout with an integrated guided mode resonance (GMR) structure, applying it to precision readout of a nanomechanical oscillator. Specifically, we fabricate subwavelength gratings into 100-nm-thick Si$_3$N$_4$ membranes and record their vibration by direct transmission measurements.   The narrow linewidth $\approx 2.5\;\text{mrad}$ of the GMR enables a shot-noise-limited displacement imprecision of $ 10^{-9}\;\t{rad/\sqrt{Hz}}$ with nanowatts of optical power, sufficient to resolve the thermal motion of a $Q\approx 10^6$ torsion mode with a signal-to-noise ratio of $47$ dB. Control experiments based on polarization and wavelength detuning confirm that the measured signal arises from GMR-mediated transduction. These results establish guided-mode resonance as an on-chip approach to angular displacement readout in quantum optomechanical sensors.
	\end{abstract}
	
	\maketitle

Precise angular displacement readout of a mechanical oscillator is a common task in science and engineering, from atomic force microscopy \cite{manalis1996high} to equivalence principle tests \cite{schlamminger2008test}. The traditional optical lever method \cite{jones2002optical_lever,manalis1996high,schlamminger2008test} offers high sensitivity and quantum efficiency \cite{pluchar2024quantum}, but relies on free-space beam propagation over extended path lengths. Next-generation optomechanical sensors demand both sensitivity and compactness, motivating investigation of photonic structures sensitive to angular displacement. Guided mode resonance (GMR) structures \cite{quaranta2018gmr_review,wang1993gmr_theory,fan2002analysis,liu1998higheffgmr} present a solution that combines sharp angular response and compatibility with nanofabrication methods used in quantum optomechanics \cite{mitra2024narrow,bui2012high,kini2020suspended}; however, to date GMR-based angle sensing has been limited to static \cite{karrock2015pressure} and low frequency \cite{mohamad2026gmrvibration} applications limited by classical noise.

Here we demonstrate use of GMR for shot-noise-limited angular displacement readout of a nanomechanical oscillator. We highlight the coherently-enhanced nature of the readout---a challenge for free-space approaches due the multimode nature of the optomechanical coupling \cite{shimoda2022coherent}---which manifests in a conventional cavity optomechanical framework \cite{aspelmeyer2014cavity} as a shot-noise-limited angular displacement imprecision of 
\begin{equation}
S_\theta^\t{imp} \gtrsim \frac{\kappa^2}{8G_\theta^2}\frac{hc}{\lambda P} = \frac{\Delta\theta^2}{8N}
\end{equation}
where $\kappa$ and $G_\theta = \partial \omega/\partial \theta$ are the linewidth and optomechanical coupling to the resonance frequency $\omega_0 =c/\lambda_0$, respectively, $\Delta \theta = \kappa/G_\theta$ is the angular resonance linewidth, and $P$ ($N=\lambda_0 P/hc$) is the detected 
optical power (photon flux).  By patterning a 25-nm-thick grating into a 100-nm-thick Si$_3$N$_4$ membrane, we realize a $\lambda_0\approx 1\;\mu\t{m}$ GMR with $\kappa/(2\pi)\sim 1\;\t{THz}$,   $G/(2\pi) \sim 100\;\t{GHz/mrad}$, and $\Delta\theta \sim 1\;\t{mrad}$, enabling readout of a higher-order membrane mode with an imprecision of $S_\theta\sim (1\;\t{{nrad}/\sqrt{Hz}})^2$ using $P\sim 100$~nW ($N \sim 10^{11}\;\t{s}^{-1}$).  We conclude by comparing the quantum limits of optical lever and GMR approaches and demonstrating thermal-noise-limited readout of a high-$Q$ torsion oscillator. 

\begin{figure}[t!]
\includegraphics[width=1.0\linewidth]{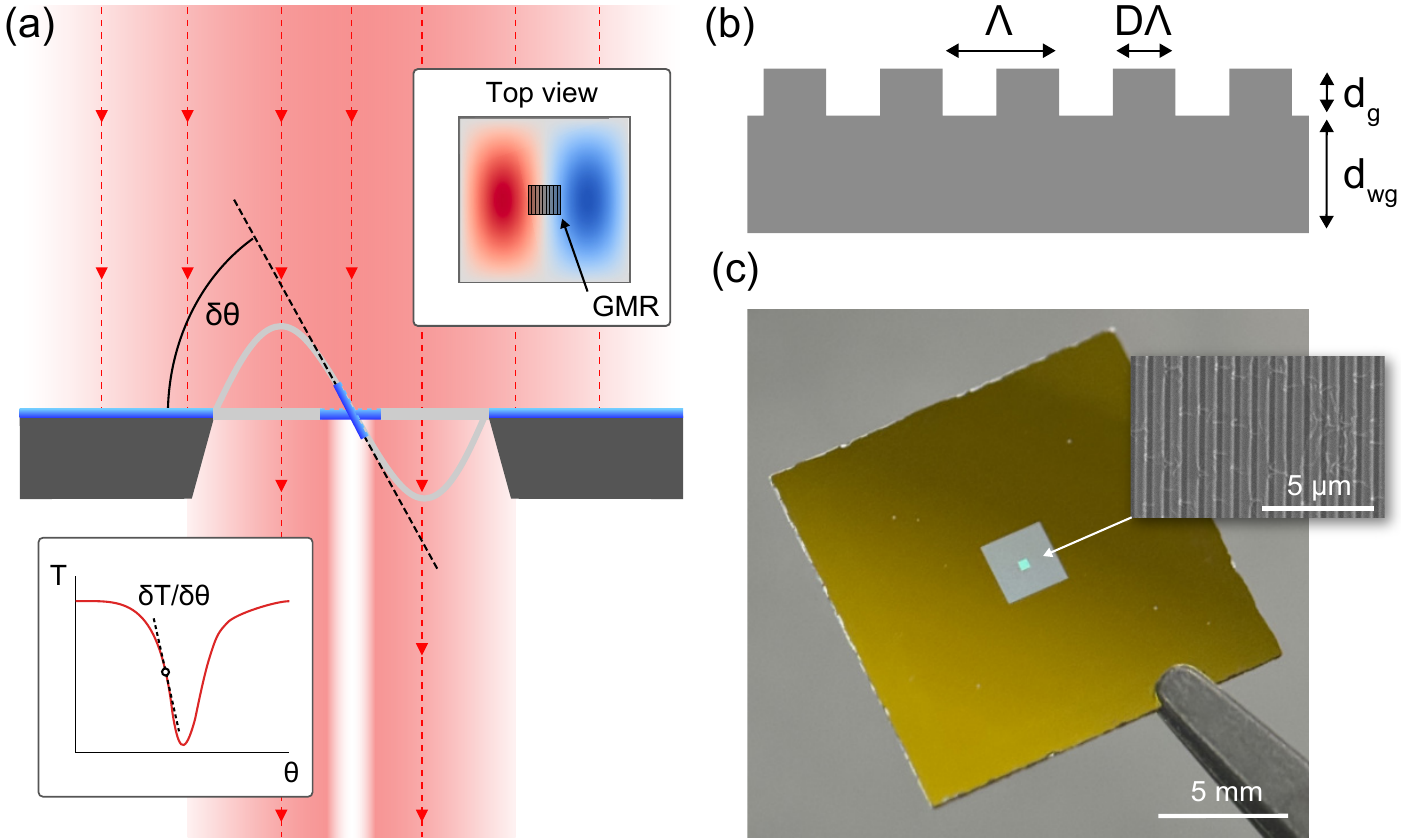}
\caption{Optomechanical coupling of a Si$_3$N$_4$ membrane to a guided mode resonance (GMR). (a) Membrane displacement modifies the local angle of incidence $\theta$ of a laser beam incident on an embedded grating slab structure, altering the transmittivity $T$ of a GMR. Inset: Orientation of the grating  relative to the mechanical mode.
    (b) Grating geometry:  period $\Lambda$, duty cycle $D$, grating thickness $d_\t{g}$, waveguide thickness $d_\t{wg}$. 
    (c) Photo of 2.5 mm Si$_3$N$_4$ membrane with embedded grating.  \mbox{Inset: scanning electron micrograph of the grating.}}
    \label{fig:figure1}
    \vspace{-2mm}
\end{figure}

\looseness=-1
\textit{Guided-mode resonance -} A brief overview of guided mode resonance (GMR) is germane to understanding our approach.  GMRs arise in subwavelength grating slabs that support leaky waveguide modes \cite{quaranta2018gmr_review,wang1993gmr_theory,fan2002analysis}. When incident light is near the resonance condition $\omega \approx \omega_0(\theta)$, it couples into the mode and interferes with the directly transmitted (or reflected) wave, resulting in a Fano-shaped transmission spectrum (Fig. \ref{fig:figure1}a inset)
\begin{equation}
T(\omega,\theta) \approx T_0\left|1+\frac{q}{1+2i(\omega-\omega_0[\theta])/\kappa}\right|^2
\end{equation}
where $T_0$ is the background transmittivity of the grating and $q$ is a (generally complex) Fano parameter. The GMR linewidth $\kappa$ is governed by structural parameters such as the grating period, waveguide and grating thickness, and refractive index contrast, all of which can be finely tuned through simulation \cite{zhou2019gmr_fom} and nanofabrication. These features have made GMRs attractive for a diversity of photonic-integrated sensing applications \cite{quaranta2018gmr_review} from biology \cite{sahoo2017gmrphase, zhou2019gmr_fom, lin2017gmr_intensity} to structural engineering~\cite{mohamad2026gmrvibration}. %effective candidate for optomechanical transduction demonstrated by work on biosensing based on detection modalities of wavelength shift, phase shift, angular shift, and intensity shift at fixed wavelength \cite{quaranta2018gmr_review, sahoo2017gmrphase, zhou2019gmr_fom, lin2017gmr_intensity}. As well, GMRs have been applied outside biosensing to detect structural vibrations  and atomic spin precessions via optical rotation \cite{sun2025gmratomicspin}.

\begin{figure*}[ht]
\includegraphics[width=\linewidth]{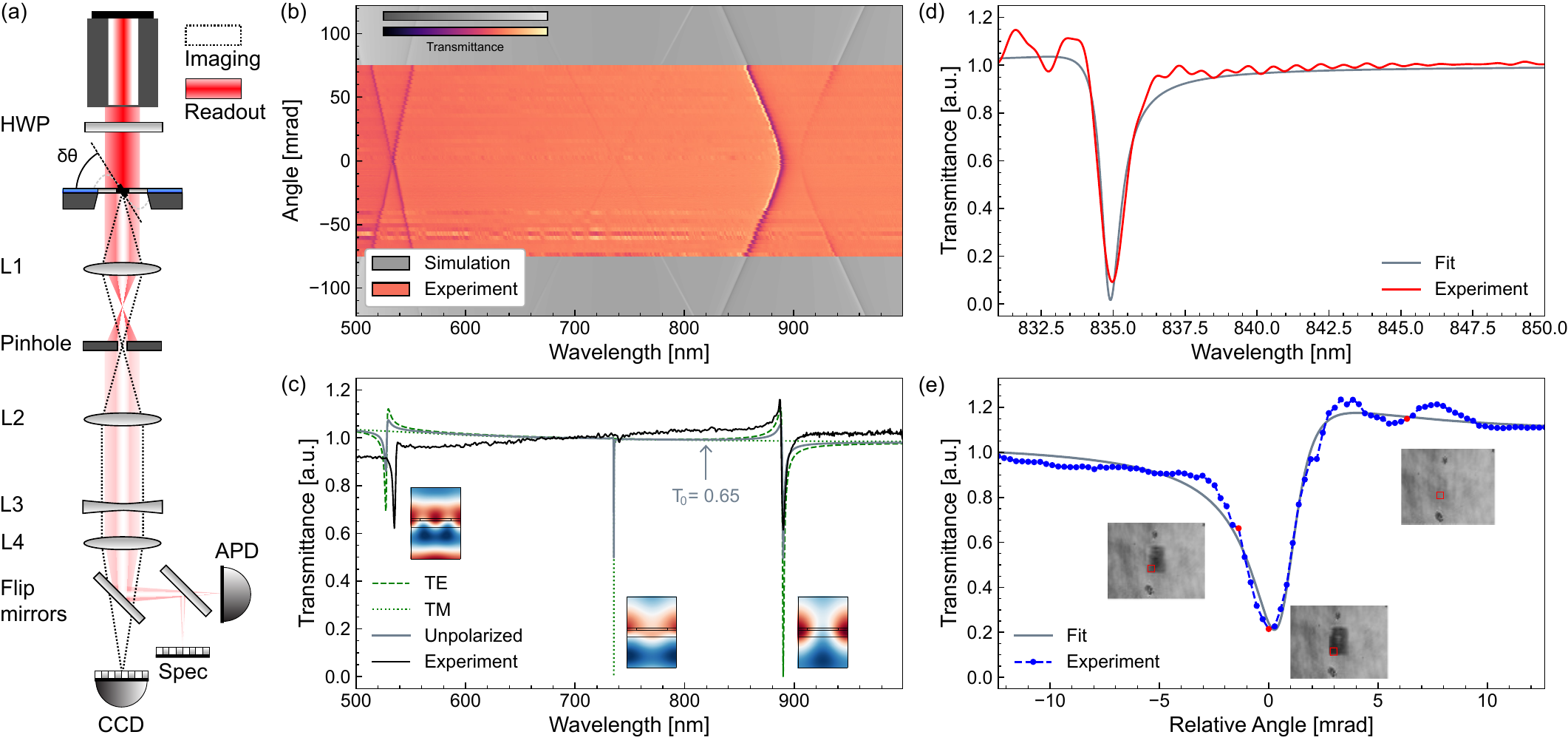}
    \caption{Optical characterization of guided mode resonance (GMR) device. (a) Experimental setup for measuring wavelength $\lambda$ and incidence-angle $\theta$ dependent membrane transmittance $T(\lambda,\theta)$ using broadband (supercontinuum) and wavelength-tunable narrowband (external cavity diode) laser light.  A half-wave plate (HWP) controls the laser polarization. Lens L1 forms a 4f system for spatial filtering with a pinhole. Lenses L2-L4 reimage the laser onto a CCD camera, spectrometer (Spec), or avalanche photodiode (APD).
    (b) Broadband measurement of $T(\lambda,  \theta)$ using the spectrometer, revealing several GMRs and avoided crossings. % are evident. %Data between dotted lines is a fine angle scan using a micrometer. 
     (c) Simulated and measured $T(\lambda,\theta)$ at normal incidence, $\theta =0$, for different polarizations. Insets: simulated GMR field profiles. 
    (d) Narrowband measurement of $T(\lambda,\theta)$ at fixed incidence angle using wavelength-scanned diode laser and CCD for readout. (e) Narrowband measurement of $T(\lambda,\theta)$ at fixed wavelength using a gimbal mount to tilt the device and CCD for readout. % used to extract slope $T'_\theta$ slope for displacement measurement calibration. 
    Insets: CCD images at highlighted points. Red boxes indicated pinhole-apertured regions.} %for displacement measurements in Fig.~\ref{fig:figure3}.}
    \label{fig:figure2}
\end{figure*}

Displacement readout of a mechanical oscillator is enabled by patterning a grating structure onto its surface and monitoring the transmitted (or reflected) field near a GMR resonance. In the limit of small displacement, the GMR can be modeled as sensitive to the local angular displacement
\begin{equation}
\theta(x,y) = \hat{h}\cdot \nabla u(x,y)
\end{equation}
where $u(x,y)$ is the tranvsere displacement profile of the oscillator and $\hat{h}$ is a unit vector normal to the grating. Monitoring a fraction of the transmitted power then yields a shot-noise-limited angular displacement sensitivity 
\begin{equation}
S_\theta^\t{imp} = \frac{2}{N}\left(\frac{T}{\partial T/\partial\theta}\right)^2 = \frac{\Delta\theta^2}{8 N}\frac{T}{\beta}
\label{eqn:gmr_angular_imp}
\end{equation}
where $\beta = (T'_\theta\Delta\theta/4)^2/T\le 1$ expresses the measurement efficiency per incident photon $N/T$, which depends on the Fano lineshape and detuning (e.g., $\beta \approx 0.42T_0$ and $\beta/T\approx 1.7$ for a Lorentzian $q = -1$  detuned to maximize $T'_\theta$.)

\begin{figure*}[ht]
  \includegraphics[width=\linewidth]{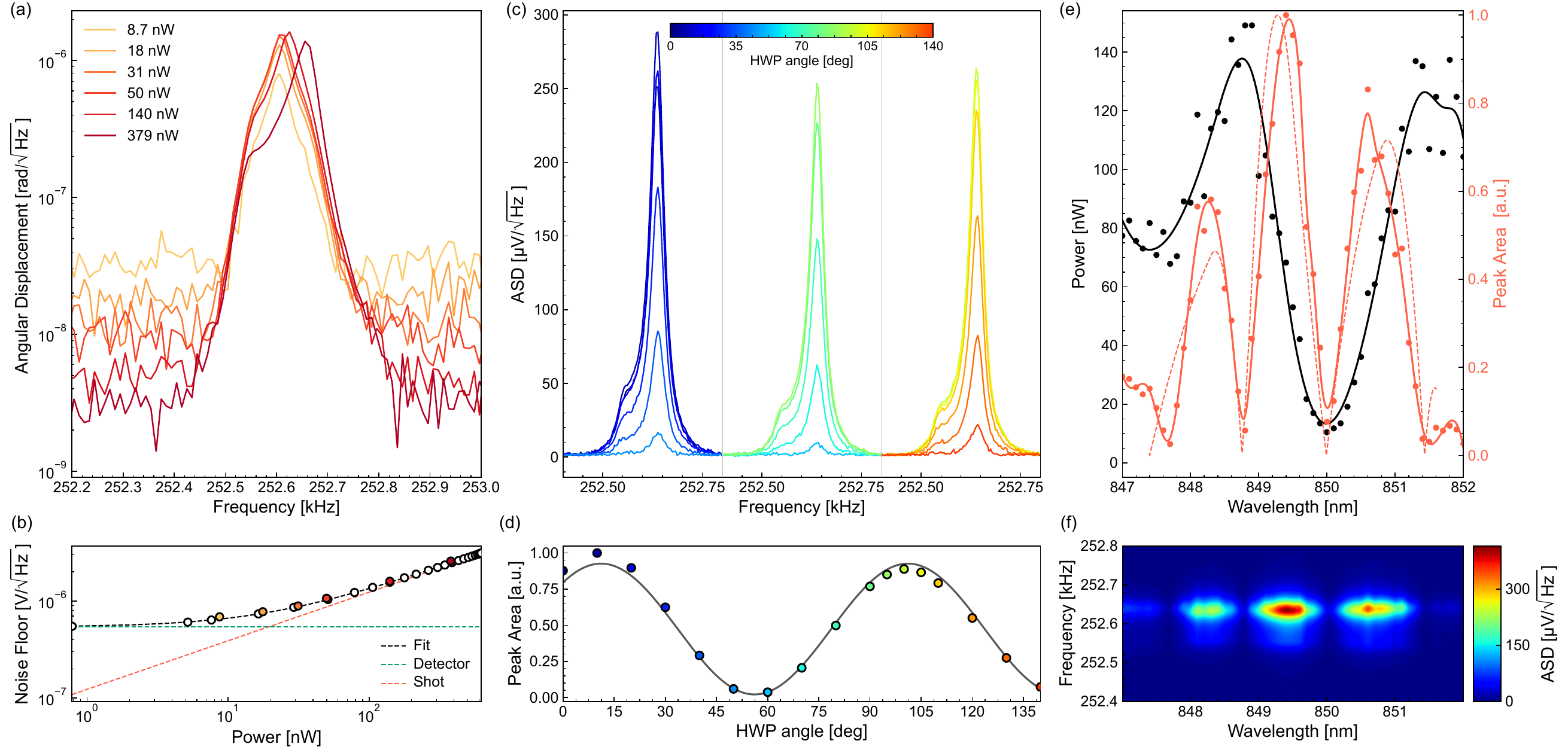}
     \caption{Angular displacement measurements and validation of optomechanical coupling. 
   (a) Driven angular displacement spectra for various detected powers $P$.  The two peaks corresponding to degenerate mechanical modes of the square membrane. (b) Noise floor in raw voltage units versus $P$, showing shot-noise scaling for $P\gtrsim 100\;\t{nW}$.
    (c) Driven angular displacement spectra for different linear polarizaion orientations of the incident field, varied with a rotatable half waveplate (HWP).  
    (d) RMS displacement versus HWP angle.
    (e) Detector optical power (black) and RMS displacement versus probe laser wavelength. 
    (f) Spectrograph of displacement spectra (panel a) versus probe laser wavelength.}
    \label{fig:figure3}
%\vspace{-1mm}
\end{figure*}

To explore these concepts, we fabricated a $290\times 290\;\mu\t{m}^2$ grating onto a 100-nm-thick, $L = 2.5\;\t{mm}$ wide square Si$_3$N$_4$ membrane, following a standard e-beam lithography and wet-release method \cite{agrawal2024focusing,hyatt2025fabrication}. Grating period $\Lambda \approx 700\;\t{nm}$, duty cycle $D \approx 5/7$, and thickness $d_\t{g} \approx 25\;\t{nm}$ were chosen according to Rigorous Coupled-Wave Analysis simulations \cite{liu2012s4} to realize a GMR near normal incidence and $\lambda_0 \approx 890\;\t{nm}$.  The grating was positioned near the membrane center $(x_0,y_0)=(L/2,L/2)$, enabling sensitivity to the antisymmetric second-order flexural mode $u(x,y) = A\sin(\pi x/L)\sin(2\pi y/L)$ with local angular displacement $\theta(x_0,y_0)\approx 2A\pi/L$.

Optical characterization of the GMR device is shown in Fig.~\ref{fig:figure2}. We first created a broadband hyperspectral map of the transmittivity $T(\lambda, \theta)$ using a collimated supercontinuum source and spectrometer, yielding qualitative agreement with model over visible wavelengths $\lambda \approx [500\;\t{nm}, 1000\;\t{nm}]$ and incicidence angles $\theta= [-4^\circ,4^\circ]$ (Fig. \ref{fig:figure2}b,c).  Higher resolution cross-sections were generated using a wavelength tunable diode laser, revealing a Fano resonance \cite{fano} with $\kappa/(2\pi) \approx 0.41$ THz (0.95 nm) and $\Delta \theta \approx 2.5\;\t{mrad}$ at $(\lambda_0,\theta_0) \approx (835\;\t{nm},6^\circ)$ (Fig. \ref{fig:figure2}d,e).  %For all measurements, with the sample chip was fixed to a calibrated gimbal mount. 
For these measurements, the laser beam was expanded to a diameter much larger than the grating, and the transmission of a patch of the grating was selected using a CCD camera as shown in the  Fig. \ref{fig:figure2}e insets.

% with the sample chip fixed to a calibrated gimbal mount
Displacement measurements were performed with the sample housed in a high-vacuum chamber ($\sim 10^{-8}\;\t{mbar}$) and the CCD replaced with a high-speed avalanche photodetector (Thorlabs APD210) and a 100-$\mu$m-wide pinhole, enabling a local patch of the membrane to be monitored.  While $Q\sim10^6$ silicon nitride membranes with embedded crystal structures have been realized and studied \cite{kemiktarak2012mechanically,bui2012high,akbar2022mechanicalpatternedmembranes}, the device for this study exhibited $Q\approx 5000$ for the targeted antisymmetric $(2,1)$ mode at $f\approx 250$ kHz (Fig. \ref{fig:figure1}a inset), making thermal noise difficult to observe.  We therefore performed driven measurements with an external piezo affixed to the vacuum chamber.% and a digital network analyzer (Zurich Instruments MFLI).

Driven displacement measurements at frequencies around the near-degenerate $(2,1)$ and $(1,2)$ membrane modes are shown in Fig. \ref{fig:figure3}a, using a digital lock-in amplifier (Zurich Instruments MFLI) to record the photodetector output $V$. To maximize sensitivity, the chip was tilted to near a resonance $(\theta_0,\lambda_0)$ at fixed wavelength; the wavelength was then tuned to the corresponding maximum slope $\partial V/\partial \lambda$ on Fig.~\ref{fig:figure2}d, which is equivalent to the maximum slope $\partial V/\partial \theta = \partial V/\partial \lambda\cdot \partial \lambda_0/\partial \theta_0$ in Fig.~\ref{fig:figure2}e, where $\partial \lambda_0/\partial\theta_0=0.39\;\t{nm/mrad}$ \cite{slope} is the slope of the resonance contour in Fig. \ref{fig:figure2}b.  The driven response was converted to PSD units $S_V$ by normalized by lock-in resolution bandwidth, and expressed in displacement units using $S_\theta = (\partial V/\partial\theta)^{-2} S_V =(\partial V/\partial \lambda\cdot \partial \lambda_0/\partial \theta_0)^{-2}S_V $.

\begin{figure}[t!]
\vspace{-1mm}
\includegraphics[width=\linewidth]{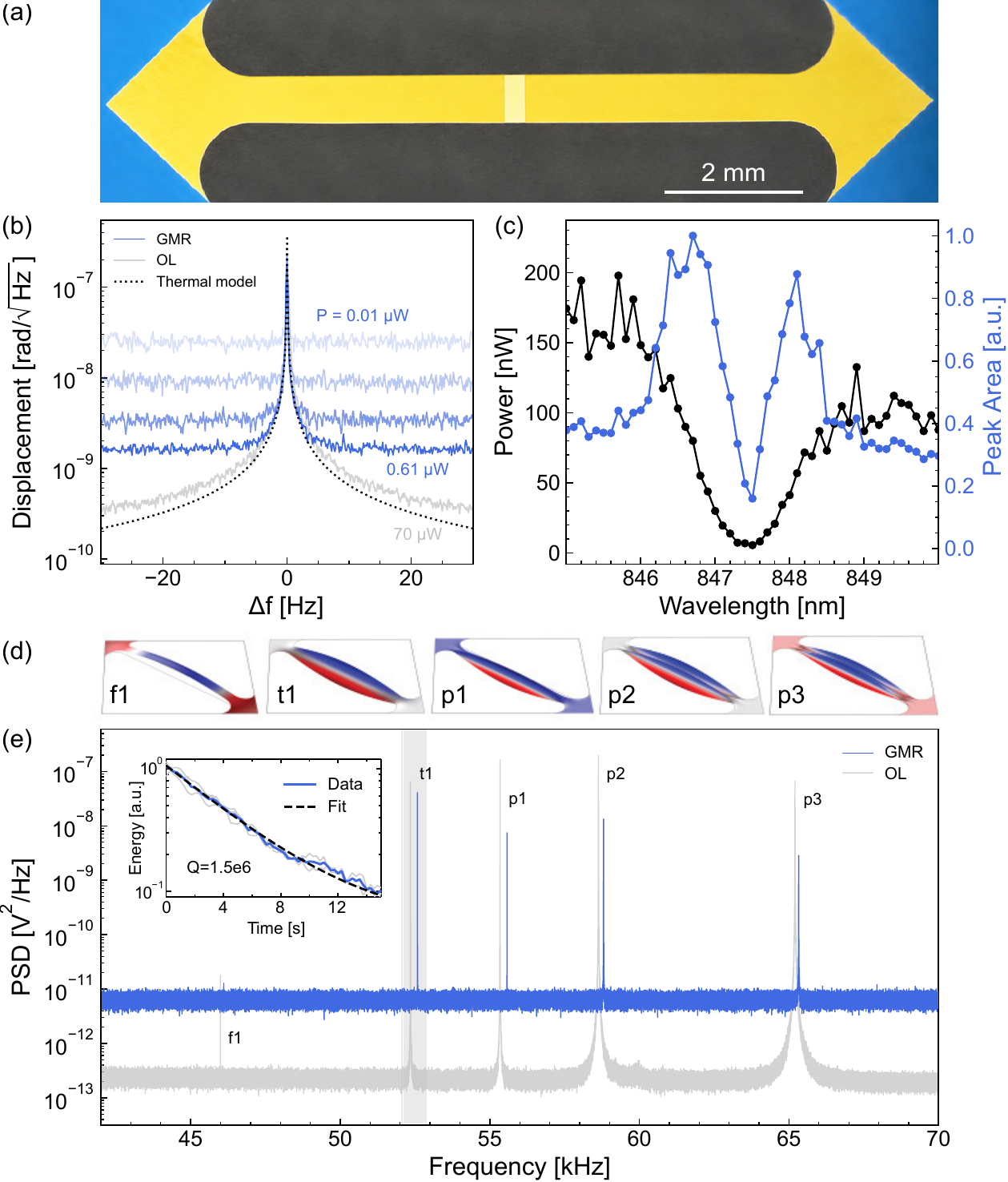}
    \caption{GMR readout of a nanomechanical torsion oscillator. (a) Microscope image of the device: a 100-nm-thick, 0.4-mm-wide, 7-mm-long Si$_3$N$_4$ nanoribbon \cite{hyatt2025bayesian} with embedded grating similar to that in Fig.~\ref{fig:figure1}. (b) Thermal-noise-limited readout of the fundamental torsion mode via GMR transmission and optical lever (OL) methods with varying detected power $P$. % = 0.24\;\mu\t{W}$, $0.54\;\mu\t{W}$ and $70\;\mu\t{W}$, respectively. 
    (c) Detected power and raw signal variance (area under thermal noise peak) versus wavelength, used to optimize the signal-to-noise ratio in (b).  (d) Simulated nanoribbon mechanicaal modeshapes. (e) Broadband measurements for the data in (b).  Inset: Ringdown measurement of torsion mode.}
    \label{fig:figure4}
    \vspace{-2mm}
\end{figure}

For detected powers from 9 nW and 380 nW, we infer an angular displacement imprecision from $30\;\t{nrad/\sqrt{Hz}}$ to $3\;\t{nrad/\sqrt{Hz}}$, respectively.  The scaling of the raw noise floor (after subtracting detector noise) is consistent with shot-noise (Fig. \ref{fig:figure3}b).  Comparing to Eq. \ref{eqn:gmr_angular_imp} with a measured linewidth  $\Delta \theta \approx 2.5\;\t{mrad}$, we infer a measurement efficiency $\beta\approx 0.05T$, $\sim3$ times smaller than predicted by the lineshape in Fig. \ref{fig:figure4}e.
%Comparing to Eq. 4 with measured linewidth $\Delta \theta \approx 2.3\;\t{mrad}$, we infer a measurement efficiency of $\beta\approx 0.2$, in good agreement with the maximum value of $\beta = 3/16$.

As further confirmation of GMR optomechanical coupling, we repeated displacement measurements as a function of polarization (Fig. \ref{fig:figure3}c,d) and wavelength (Fig. \ref{fig:figure3}e,f) for the nominal operating point in Fig. \ref{fig:figure3}a.  As anticipated, sensitivity was found to depend sinusoidally on input linear polarization angle, vanishing at an angle normal to the grating, where the GMR contrast is minimal.  Tuning the laser wavelength likewise samples different local gradients of the 2D GMR resonance, $T(\lambda,\theta)$.  For a fixed $\theta$, the sensitivity $T'_\theta $ is expected to vanish near maxima of the wavelength gradient $T'_\lambda = 0$.  This is corroborated by cross-sections of the driven displacement spectrograph $S_\theta(\omega,\lambda)$ shown in Fig. \ref{fig:figure3}e,f.

\textit{Comparison to an optical lever-} Our results show that a GMR structure can serve as an integrated cavity-enhanced angular displacement sensor.  It is interesting to compare its performance to that of an optical lever (OL), whose shot-noise-limited displacement sensitivity can be expressed as \cite{pluchar2024quantum}
\begin{equation}
S_\theta^\t{imp} \ge \frac{\theta_\t{D}^2}{8N}\frac{1}{\beta_\t{OL}}
\end{equation}
where $\theta_\t{D} = \lambda/(\pi w_0)$ is the diffraction angle of the reflected laser and $\beta_\t{OL}\le 1$ is the measurement efficiency (for a split photodiode, $\beta_\t{OL} \le 2/\pi$ \cite{pluchar2024quantum}).  %Combined with Eq. 4, the GMR sensitivity can be recast as an effective OL spot size
Comparing to Eq. \ref{eqn:gmr_angular_imp} at equal incident power and background reflectivity $R_0 = 1-T_0$, the GMR sensitivity can be recast as an effective OL spot size
\begin{equation}
w_{0} = \frac{\lambda}{\pi\Delta \theta}\sqrt{\frac{\beta}{\beta_\t{OL} R_0}}\approx\frac{\lambda}{4\Delta\theta}\sqrt{\frac{\beta/T_0}{0.42}\frac{2/\pi}{\beta_\t{OL}}\frac{T_0}{R_0}}
\end{equation}
% Comparing to Eq. 4 with a common incident power and background reflectivity $R_0 = 1-T_0$, the GMR sensitivity can be recast as an effective OL spot size
% \begin{equation}
% w_{0,\t{GMR}} = \frac{\lambda}{\pi\Delta \theta}\sqrt{\frac{\beta_\t{GMR}}{\beta_\t{OL} R_0}}\approx\frac{\lambda}{2\Delta\theta}\sqrt{\frac{\beta}{0.42}\frac{2/\pi}{\beta_\t{OL}}\frac{1}{R_0 T_0}}
% \end{equation}
% \begin{equation}
% w_0 = \frac{\lambda}{\pi\Delta \theta}\sqrt{\frac{\beta/T}{\beta_\t{OL}}}\approx\frac{\lambda}{2\Delta\theta}\sqrt{\frac{\beta/T}{1.7}\frac{2/\pi}{\beta_\t{OL}}}
% \end{equation}
which for $\Delta \theta\approx 2.5\;\t{mrad}$ corresponds to $\lambda/(4\Delta\theta) \approx 85\;\mu\t{m}$.  Recent studies of quantum-limited OL measurements on nanomechanical oscillators \cite{hao2024back,pluchar2024quantum,shin2024laser} have employed similar spot sizes and realized similar measurement efficiences $\beta_\t{OL}\approx 0.5$. A distinction of the GMR is that it is a local displacement probe.  This may have significant advantages at low frequencies where apparatus and beam pointing noise are known to limit the sensitivity of OL measurements \cite{shin2024laser}.  It can also enable robust free-space readout of chipscale sensors, by replacing sensitive alignment with wavelength tuning.  

\textit{Displacement sensitivity limits-} 
In our experiment, the achieved displacement imprecision  $S_\theta \approx (3\;\t{nrad/\sqrt{Hz}})^2$ was limited by the throughput of our plane-wave imaging scheme, which allowed only $\eta \sim 0.01\%$ of the optical power incident on the membrane ($\sim 1\,\t{mW}$) to reach the photodector. In principle $\eta\sim 1$ is achievable with an aperture-filling laser beam $2w_0\approx L_\t{GMR}$ \cite{toftvandborg2021collimation}, implying that $S_\theta^\t{imp}<(0.1\;\t{nrad/\sqrt{Hz}})^2$ is accessible with our current GMR linewidth $\Delta \theta\approx 2.5\;\t{mrad}$ and probe power. Achieving greater precision requires decreasing GMR linewidth and, or, increasing power. A practical lower bound is given by the diffraction limit $\Delta\theta \gtrsim\lambda/L_\t{GMR}$ \cite{toftvandborg2021collimation,bendickson2001guided}, implying a $L_\t{GMR} = 1\;\t{mm}$ grating probed with $P= 10\;\t{mW}$ should enable $S_\theta^\t{imp}<(0.01\;\t{nrad/\sqrt{Hz}})^2$, on par with the state-of-the-art OL measurements \cite{hao2024back,shin2024laser,pluchar2024quantum}.

\textit{Nanomechanical torque sensing-} A promising application of GMR-based angular displacement sensing is readout of a nanomechanical torsion oscillator for precision torque sensing \cite{davis2010nanotorsional,zhang2013nanomechanical}.  Recently, centimeter-scale Si$_3$N$_4$ nanoribbons have been reported with ultrahigh quality factors $Q \sim 10^8$ at acoustic frequences $f_0\sim 100\;\t{kHz}$, yielding thermal torque sensitivities of $S_\tau^\t{th}\sim (10\;\t{zNm/\sqrt{Hz}})^2$ and resonant thermal displacements of $S_\theta^\t{th}(f_0)\sim (\mu\t{rad/\sqrt{Hz}})^2$ \cite{pratt2023nanoscale,hyatt2025bayesian}.  Embedding a GMR into such a nanoribbon could enable $\t{zNm/\sqrt{Hz}}$ torque sensitivity with picowatts of optical power. We present progress towards this goal in Fig. \ref{fig:figure4}, by fabricating a grating similar to that in Fig.~\ref{fig:figure1} into a $w = 400\;\mu\t{m}$-wide, $h = 100\;\t{nm}$-thick, 7-mm-long Si$_3$N$_4$ nanoribbon~\cite{hyatt2025bayesian}.  Broadband displacement measurements shown in Fig. \ref{fig:figure4}b were made by probing the side of a GMR with $P \approx 0.6\;\mu\t{W}$, revealing a series of thermal noise peaks, including a $Q\approx 1.5\times 10^6$ noise peak at $f_0 = 52.5\,\t{kHz}$, corresponding to the fundamental torsion mode. Bootstrapping a Lorentzian fit to the simulated peak thermal displacement $S_\theta^\t{th}(f_0) \approx (0.37\,\mu\t{rad}/\sqrt{\t{Hz}})^2$, we infer a readout imprecision of {$S_\theta^\t{imp} \approx (1.6\, \t{nrad/\sqrt{Hz}})^2$, qualitatively consistent with the direct calibration in Fig.~\ref{fig:figure3}a. Independent of calibration, the high signal-to-noise ratio $S_\theta^\t{th}[f_0]/S_\theta^\t{imp}\approx 47$ dB
corresponds to realizing a thermal torque sensitivity of $S_\tau^\t{th}\approx (0.1\;\t{aNm/\sqrt{Hz}})^2$ \cite{hyatt2025bayesian} over a bandwidth of $\Delta f = (f_0/Q)(S_\theta^\t{th}[f_0]/S_\theta^\t{imp})^{1/2} \approx 8\;\t{Hz}$.

 \textit{Summary \& outlook -} We have explored optomechanical coupling to a guided mode resonance (GMR) as a method for coherently-enhanced angular displacement readout of a mechanical oscillator \cite{shimoda2022coherent},  comparing its quantum limits to that of the optical lever \cite{pluchar2024quantum}, and conducting proof-of-principle experiments with Si$_3$N$_4$ membranes patterned with subwavelength gratings.  Using direct transmission measurementes, we realized $\t{nrad/\sqrt{Hz}}$ sensitivity with nanowatts of optical power, limited by a diffraction-limited angular GMR linewidth $\Delta\theta\approx 2.5\,\t{mrad}$, and corresponding to a $\mathcal{O}(1)$ quantum efficiency.  We applied this displacement sensitivity to readout of a high-$Q$ Si$_3$N$_4$ torsion nanoribbon, resolving thermal noise with a $47$ dB signal-to-noise ratio, corresponding to a $ 0.1\;\t{aNm}/\sqrt{\t{Hz}}$ torque sensitivity over a $8\;\t{Hz}$ bandwidth.%with subwavelength gratings.

Looking forward, combining low loss and lithographic control over the optomechanical properties of suspended GMR structures offers diverse opportunities.  Hybrid GMR-Fabry-Perot cavities have been widely explored \cite{vcernotik2019cavity,fitzgerald2021cavity} and offer a route to enhanced optomechanical interactions \cite{monsel2024dissipative,mitra2026cavity}. Our focus has been on direct GMR optomechanical coupling to angular displacement, reducing engineering complexity and avoiding constraints posed by the nondegeneracy of Fabry–Perot spatial modes \cite{shimoda2022coherent}. A promising application is optomechanical inertial sensing \cite{liu2021progress}, which requires integrated displacement readout methods that are both sensitive and robust to parasitic vibrations, and for which chip-scale torsion oscillators have emerged as an attractive platform \cite{condos2025ultralow}.
GMR optomechanical coupling is also not limited to angular displacement~\cite{zanotto2022optomechanical}. Matching the local grating orientation to the gradient of the mechanical modeshape can enable selective coupling to specific mechanical modes, as well as tailored intra- and intermodal coupling between tranverse spatial modes of the optical field \cite{pluchar2025imaging}.  In conjunction with high-$Q$ nanomechanics \cite{engelsen2024ultrahigh}, this approach could provide a general framework for resonant spatial-mode optomechanics.

\color{black}

%\vspace{5mm}
\section*{Acknowledgements}
The authors thank Atkin Hyatt and Justin Murante for helpful discussions and experimental assistance. This work was supported by the National Science Foundation (NSF), award no. 2209473 and by the Office of Naval Research MURI, award no. N000142612102. D.T.B. acknowledges additional support from the Friends of Tucson Optics scholarship. O.E.A. acknowledges support from the Secretaría de Ciencia, Humanidades, Tecnología e Innovación (SECIHTI), Mexico (CVU 2023435). The reactive ion etcher used for this study was funded by an NSF MRI grant, ECCS-1725571.

\bibliography{ref}

\begin{thebibliography}{42}
\bibitem [{slo()}]{slope}%
  \BibitemOpen
  \href@noop {} {\bibinfo {title} {As a consistency check, we note that the ratio of the linewidths and maximum slopes in {F}igs. 2e and 2d is ${\Delta}\lambda/{\Delta}\theta\approx 0.38$ nm/mrad, and ${T}'_\theta/{T}'_\lambda \approx 0.41$ nm/mrad, respectively.}}\BibitemShut {Stop}%
\end{thebibliography}

\end{document}

\raggedbottom

\title{Supplemental Information for ``Angular displacement readout of a mechanical oscillator with a guided mode resonance''}

	\author{D.   Torres-Barajas}
%	\email{dtorresb@arizona.edu}
	\affiliation{Wyant College of Optical Sciences, University of Arizona, Tucson, AZ 85721, USA}

    \author{O. Angulo}
	\affiliation{Wyant College of Optical Sciences, University of Arizona, Tucson, AZ 85721, USA}

	\author{A. R. Agrawal}
	\affiliation{Wyant College of Optical Sciences, University of Arizona, Tucson, AZ 85721, USA}

  %  \author{A. D. Hyatt}
%	\affiliation{Wyant College of Optical Sciences, University of Arizona, Tucson, AZ 85721, USA}
    
	\author{M. ElKabbash}
	\affiliation{Wyant College of Optical Sciences, University of Arizona, Tucson, AZ 85721, USA}

    \author{D. J. Wilson}
%	\email{dalziel@arizona.edu}
	\affiliation{Wyant College of Optical Sciences, University of Arizona, Tucson, AZ 85721, USA}
    
\begin{abstract}
Here we provide details about the method used to a fabricate Si$_3$N$_4$ nanoribbon with embedded subwavelength grating shown in Fig. 4 of the main text.
\end{abstract}

\maketitle

\section{Fabrication of Si$_3$N$_4$ membranes with embedded gratings}

\begin{figure}[h!]
\includegraphics[width=0.5\linewidth]{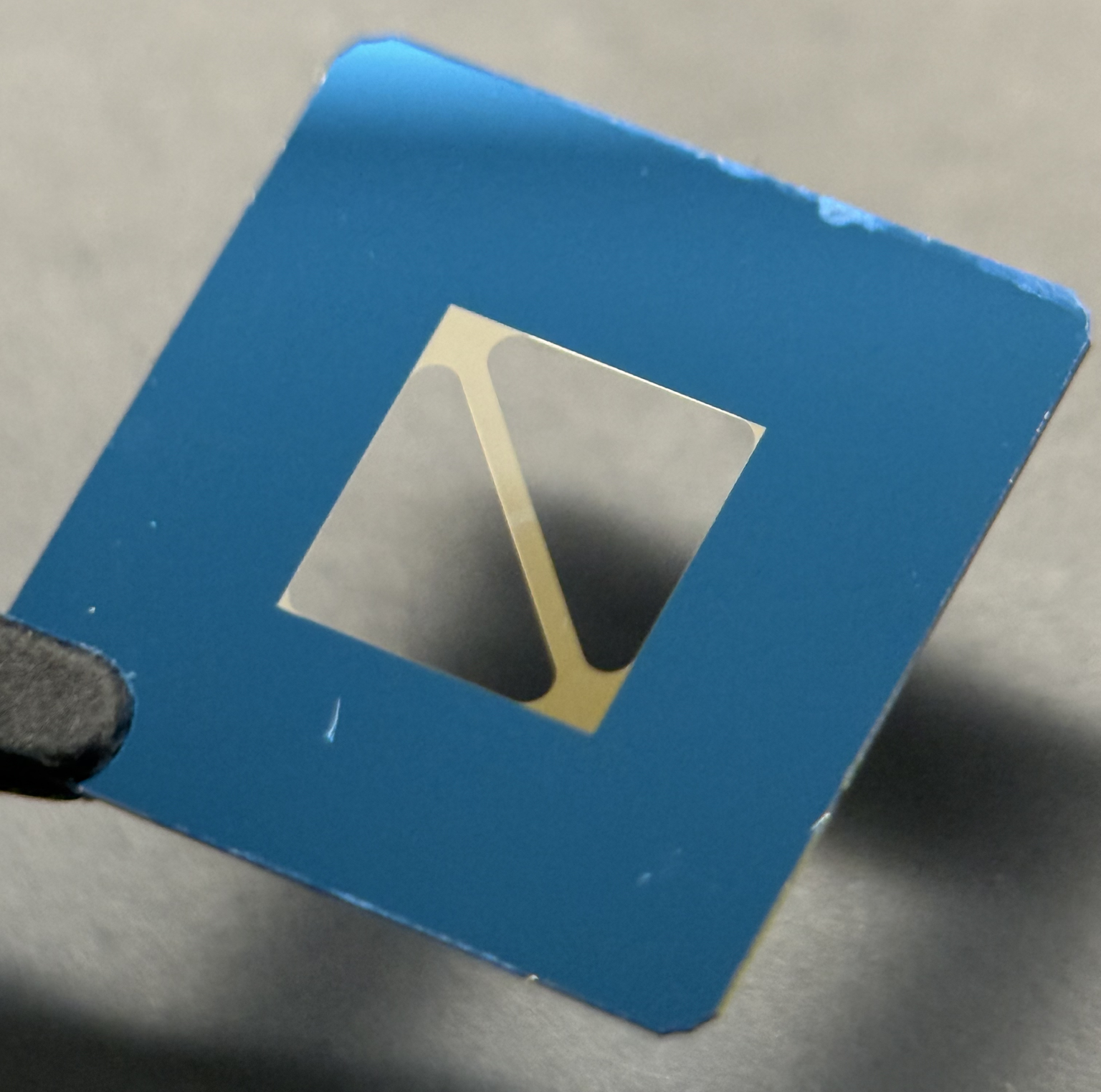}
    \caption{Photograph of Si$_3$N$_4$ nanoribbon with embedded subwavelength grating after KOH etching.}
    \label{fig:figureS1}
 %   \vspace{5mm}
\end{figure}

A photograph of the device described in  Fig. 4 of the main text---a Si$_3$N$_4$ nanoribbon with embedded subwavelength grating---is shown in Fig. \ref{fig:figureS1}.  The process flow for fabrication of this device is shown in Fig. \ref{fig:figureS2}, and is identical to that used for the pattered membrane described in main text Figs. 1-3. Details are provided below.

\subsection{Patterning of subwavelength grating}

The fabrication procedure starts with a 100-nm-thick Si$_3$N$_4$ membrane or nanoribbon prepatterned on a 200-$\t{\mu m}$-thick Si chip with a backwindow (Fig.~\ref{fig:figureS2}a), following the procedure described in~\cite{hyatt2025fabrication}. The grating is patterned using electron beam lithography (EBL, Elionix ELS-7000)%, we pattern a grating with $325~\t{nm}$ wide features and a fill factor of $0.5$. The pattern is then
and exposed using ZEP520A as a resist  (Fig.~\ref{fig:figureS2}b). For alignment of the grating pattern on the membrane, the stage coordinates of the corners of the chip were used (found using the SEM function of the EBL machine). Exposure parameters are summarized in Table~1.
\vspace{5mm}

\noindent\begin{minipage}{\columnwidth}
\begin{center}
{\small Table 1. Electron-beam exposure parameters.}\\[0.4ex]
\begin{ruledtabular}
\begin{tabular}{llll}
Beam current & $300~\t{pA}$ & Dots per field & $240000$ \\
Aperture & $60~\t{\mu m}$ & Dose & $390~\t{\mu C/cm^2}$ \\
Field size & $600~\t{\mu m}$ & Resist thickness & $400~\t{nm}$\\
\end{tabular}
\end{ruledtabular}
\end{center}
\end{minipage}

%\newpage

\subsection{Pattern transfer and release}

\begin{figure}[h!]
\includegraphics[width=0.75\linewidth]{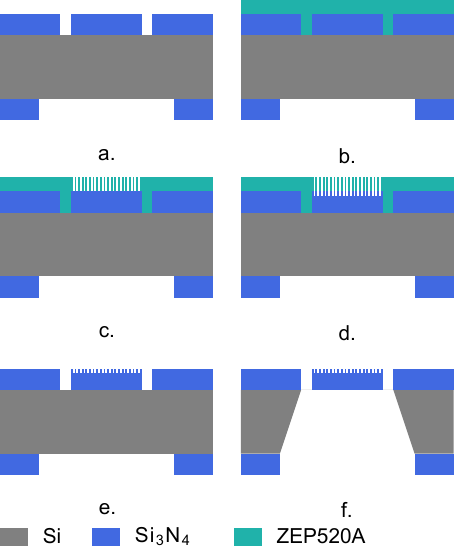}
    \caption{Process flow. (a) Starting chip. (b) Spin-coated electron beam resist. (c) Written grating pattern on the resist. (c) Pattern is transferred to the Si$_3$N$_4$ by dry etching. (e) Resist residue is cleaned. (f) Device is released by wet etching.}
    \label{fig:figureS2}
  %  \vspace{-2mm}
\end{figure}

After the exposure, the pattern was developed by immersion in Amyl Acetate for 3 minutes, followed by an IPA rinse (Fig.~\ref{fig:figureS2}c). The pattern is transferred to the Si$_3$N$_4$ using RIE with CHF$_3$, in this case we etched $20~nm$ deep into the Si$_3$N$_4$ film (Fig.~\ref{fig:figureS2}d). The e-beam resist is then cleaned using 2 consecutive immersions in NMP at $80^{\circ}$C for 5 min each, followed by water and IPA rinses (Fig.~\ref{fig:figureS2}e). The release is then performed in a $30\%$ KOH solution at $80^{\circ}C$ (Fig.~\ref{fig:figureS2}f).

\bibliography{ref}

%\vfill\null